\documentclass[aps,prb,article,twocolumn,floatfix, superscriptaddress,longbibliography]{revtex4-2}
\usepackage{lipsum}
\usepackage{graphicx}
\usepackage{epstopdf}
\usepackage{dcolumn}
\usepackage{bm}
\usepackage{physics}
\usepackage{xcolor}
\usepackage{amssymb}
\usepackage{amsmath}
\usepackage{bbold}
\usepackage{soul}
\usepackage[colorlinks=true, linkcolor=blue, urlcolor=blue, citecolor=blue]{hyperref}
\usepackage{float}

\newcommand{\affFUW}{Faculty of Physics, University of Warsaw, Pasteura 5, 02-093 Warsaw, Poland}
\newcommand{\affCent}{Centre for Quantum Optical Technologies, Centre of New Technologies, University of Warsaw, Banacha 2c, 02-097 Warsaw, Poland}
\begin{document}

\preprint{APS/123-QED}

\title{Variational non-gaussian approach to interacting spin-boson models}
    
\author{J. P. Mendon\c{c}a} 
\email{jpedromend@gmail.com}
\affiliation{\affFUW}
\affiliation{\affCent}
\author{Y. Wang}
\affiliation{Department of Chemistry, Emory University, Atlanta, Georgia 30322, USA}
\author{K. Jachymski}
\affiliation{\affFUW}

\date{\today}

\begin{abstract}
    We apply a hybrid variational framework to interacting spin-boson Hamiltonians, targeting regimes where simulations are limited by the unbounded bosonic Hilbert space and strong many-body correlations. The bosonic sector and spin-boson correlations are captured within a compact non-Gaussian variational manifold, while the minimized spin sector is obtained as the solution to an effective spin Hamiltonian. Minimization is carried out inside a self-consistent energy-minimization loop, where variational parameters are minimized and the effective Hamiltonian is solved via DMRG. The results are obtained without eliminating or truncating the photonic field. We benchmark the method on the Dicke and Dicke-Ising models by comparison to converged spin-boson DMRG, finding accurate ground-state solutions with reduced bond dimension.
\end{abstract}

\maketitle

\section{Introduction}

Spin-boson models, where discrete quantum degrees of freedom interact with one or more bosonic modes, form a central class of models in contemporary quantum science~\cite{leggett1987dynamics}. A prominent realization is cavity QED~\cite{schlawin2022cavity,basov2025review}, where atoms or artificial emitters couple coherently to a quantized electromagnetic mode, as described by seminal Rabi~\cite{rabi1936on}, Jaynes-Cummings~\cite{jaynes-cummings1963comparison}, and Dicke models~\cite{dicke1954coherence}, possibly reaching the ultrastrong coupling regime~\cite{diaz2019ultrastrong,kockum2019ultrastrong}. In the same spirit, circuit QED provides a highly successful and experimentally mature implementation of cQED with superconducting qubits coupled to microwave resonators~\cite{blais2004cavity,schoelkopf2008wiring,blais2021circuit}.
Another important example is provided by spin-phonon systems, where spin degrees of freedom couple to quantized vibrations~\cite{cross1979new,hase1993observation,kiryukhin1995direct,moseley2018spin} or to engineered vibrational modes~\cite{bennett2013phonon,li2020enhancing}, offering a complementary route to spin-boson physics in which phonons replace photons as the mediating bosonic field.
Together, these paradigmatic realizations have made spin-boson Hamiltonians a widely used effective description not only for exploring quantum many-body phenomena, but also for enabling practical directions such as quantum simulation~\cite{safavi2018verification,peraca2024quantum}, collective quantum sensing and metrology~\cite{colombo2022time,Franke2023quantum}, quantum batteries~\cite{ferraro2018high,campaioli2024colloquium,quach2022super}, and ultrafast spectroscopy~\cite{otterpohl2024coherent}.

Coupling an interacting spin subsystem to a bosonic sector results in a nontrivial interplay between the two.
In regimes where the bosons can be integrated out, this interplay is commonly captured by an effective description in which the bosonic degrees of freedom act primarily as mediators of interactions and can be eliminated from the model~\cite{porras2004effective,caballerobenitez2015quantum,romanroche2022effective,granet2023exact}.
This regime is highly accessible experimentally and well characterized in several architectures~\cite{majer2007cavity,kim2009entanglement,baumann2010dicke,ritsch2013cold,landig2016quantum,periwal2021programmable}.
Beyond the elimination limit, retaining the bosons explicitly can be essential even for the ground state properties, since spin-boson correlations and fluctuations can reshape the quantum phases and critical behavior in ways that simplified description may miss~\cite{bezvershenko2021dicke,otterpohl2022hidden,orso2025self,tolle2025fluctuation}.
Describing these systems is theoretically challenging, as energy scales need not separate cleanly, the bosonic Hilbert space is infinite-dimensional, and correlations may be strong, requiring the use of accurate non-perturbative methods.
Recent advances include matrix product states (MPS)~\cite{wall2016simulating}, digital and analog quantum-simulation~\cite{shapiro2025digital,leppakangas2018quantum}, quantum Monte Carlo (QMC) algorithms ~\cite{weber2022quantum,langheld2025quantum,dong2025quantum}, variational non-Gaussian ans\"atze~\cite{knorzer2022spin}, and semi-analytic solutions via compact dressed representations~\cite{garwola2025compact}. Together, these works underscore both the rapid methodological progress in the field and the continuing need for robust numerical methods that can accurately access strongly correlated spin-boson physics.

In this paper, we describe in detail the hybrid variational framework used in Ref.~\onlinecite{mendonca2025role} in which the bosonic degrees of freedom and spin-boson correlations are represented within a non-Gaussian variational manifold, while the spin sector is treated with DMRG/MPS inside a self-consistent energy-minimization loop. This formulation offers a distinct algorithmic pathway compared to tensor network methods based on local basis optimization (LBO)~\cite{Zhang1998LBO}. Whereas LBO addresses the infinite Hilbert space by constructing an adaptable local basis that must be updated and rotated during the DMRG sweeps, our approach decouples the bosonic optimization from the MPS calculations. By capturing the photonic mode(s) through the variational parameters of the ansatz (displacement, squeezing, and dressing), we effectively map the problem onto a self-consistent spin model. This allows the DMRG solver to focus exclusively on the spin correlations without the computational overhead of iterative basis reconfiguration. In order to provide a use case example, we apply it to the Dicke-Ising model, which has been extensively studied using a variety of methods~\cite{Rohn2020,gammelmark2011phase,langheld2025quantum,romanroche2022effective,schellenberger2026infinity,leibig2026quantitative,koziol2026}, and is known to feature several quantum phases with both first- and second-order transitions.

The remainder of the paper is organized as follows. In~\ref{sec:models} we discuss the class of models described by our framework. Section~\ref{sec:method} provides a detailed description of the method, Sec.~\ref{sec:results} contains the analysis of the Dicke-Ising model which we use as a benchmark case study, and in Sec.~\ref{sec:benchmark} the method performance is discussed. We conclude the work in Sec.~\ref{sec:conclusion}. The Appendices provide additional details and a possible extension scheme.

\section{Spin-boson models}
\label{sec:models}

We consider an ensemble of $N$ interacting two-level systems (TLS) coupled to an $M$-mode bosonic field. 
Throughout this work we represent the TLS as spin-$1/2$ systems, and we use the computational product basis of $S^z$ eigenstates for the spin sector.
The general Hamiltonian we target is
\begin{align}\label{eq:fullH}
H &= \sum_{m=1}^{M} \omega_{m} a_{m}^\dagger a_{m}
+
\sum_{i=1}^{N} \varepsilon_{i} s_{i}^{z}
+
\sum_{i=1}^{N} \sum_{m=1}^{M} g_{i m} s^\mu_{i} (a_{m} + a_{m}^\dagger )
\nonumber\\
&-
\sum_{i < j} \sum_{\gamma\in{x,y,z}} J_{i j}^{\gamma} s_{i}^{\gamma} s_{j}^{\gamma}.
\end{align}
Here, $s_i^\alpha = \sigma_i^\alpha/2$ are local spin operators, where $a_m$ ($a_m^\dagger$) annihilates (creates) a boson in mode $m$, and fixed $\mu=x$, $y$, or $z$. The parameters $\omega_m$ and $\varepsilon_i$ set the bosonic and spin energy scales, respectively, while $g_{im}$ and $J_{ij}^\gamma$ denote spin-boson couplings and spin-spin interactions, which in principle may be completely inhomogeneous, incorporate disorder, or include long-range terms. 
We introduce below some common examples that belong to this class of models.

\subsection{Dicke-type spin-boson interactions}
The Dicke model, describing an ensemble of two-level systems coupled collectively to a single photonic mode ($M=1$), stands as the paradigmatic example of a spin-boson system~\cite{dicke1954coherence}. Due to its high degree of symmetry and the presence of only one bosonic mode, the Dicke model has been systematically investigate both analytically and numerically (cf. review in Ref.~\onlinecite{Kirton2019}). In particular, it is known that mean-field theory becomes exact in the thermodynamic limit, and in the strong-coupling regime the ground state is well approximated by a product state composed of a coherent state for the photons and a spin-polarized state. We use the following normalization convention $g_{i1}\equiv 2g/\sqrt{N}$, which ensures a well-defined thermodynamic limit~\cite{hepplieb1973equilibrium,hepplieb1973on,wang1973phase}. The Hamiltonian then becomes
\begin{align}\label{eq:dicke}
    H_{\rm Dicke} &= \omega \left( a^\dagger a + \frac{1}{2} \right) + \varepsilon \sum_j s^z_j + \frac{2g}{\sqrt{N}} \sum_j s^x_j (a + a^\dagger ) \nonumber\\ 
    &= \frac{\omega}{2} (x^2 + p^2) + \varepsilon S^z + g' S^x x,
\end{align}
where we defined the collective spin operators $S^\alpha = \sum_j s^\alpha_j$, the photon quadratures $x = (a^\dagger + a)/\sqrt{2}$ and $p = i(a^\dagger - a)/\sqrt{2}$, and the scaled coupling $g' = 2g/\sqrt{N/2}$. Direct spin-spin interactions are absent $J_{ij}^\gamma=0$. Throughout this paper, we fix $\omega=\varepsilon=1$ unless stated otherwise.

The fundamental feature of the Dicke model is the prediction of the superradiant quantum phase transition (QPT). However, the validity of this equilibrium transition in atomic systems is restricted by gauge invariance arguments~\cite{rzazewski1975phase}. The standard Dicke Hamiltonian neglects the diamagnetic $\mathbf{A}^2$ term inherent to the minimal coupling scheme. When this term is included, the Thomas-Reiche-Kuhn (TRK) sum rule ensures that the diamagnetic energy contribution compensates for the mode softening induced by the paramagnetic interaction, thereby forbidding the superradiant QPT in atomic systems governed by electric dipole interactions. Such an argument was later generalized to what is today called the no-go theorem of superradiance~\cite{birula1979nogo,gawedzki1981nogo}.
However, the restrictions of the no-go theorem can be bypassed in several settings, namely driven-dissipative cavity QED schemes where the Dicke Hamiltonian emerges as an effective description of Raman transitions~\cite{dimer2007proposed}; in superconducting circuits the constraints of the TRK sum rule can be violated due to the specific topology of capacitive coupling, permitting the equilibrium superradiant phase transition~\cite{nataf2010nogo}; in magnonic crystals, the Dicke model can be realized through spin-magnon exchange interactions where the diamagnetic term is naturally absent, allowing for the observation of the superradiant phase transition in thermal equilibrium~\cite{kim2025observation}. It has also been argued that the problem is an artifact of single-mode truncation in the Coulomb gauge, and that the Dicke model remains a valid effective description in the electric-dipole gauge where the superradiant transition is physically permitted~\cite{vukics2012adequacy}.

\subsection{Extensions to the Dicke framework}
While the standard Dicke model assumes an ensemble of non-interacting spins coupled solely through the bosonic mode, experimental realizations often exhibit direct spin-spin interactions. We refer to the inclusion of such terms as extended Dicke models. 
State-of-the-art quantum simulators allow for the engineering of the spin sector with high precision. As detailed in Refs.~\cite{nguyen2018towards,monroe2021programmable}, platforms such as circular Rydberg atoms and trapped ions now allow for the engineering of programmable spin-spin couplings with highly tunable range, anisotropy, and geometry, effectively bridging the physics of collective cavity QED with that of interacting spin lattices. Alternatively, strong competing spin-spin and spin-photon interactions can be obtained in solid-state systems, specifically through the quantum simulation of extended Dicke models in magnonic materials~\cite{peraca2024quantum}. Below we show two examples, Dicke-Ising model, which was extensively studied in the literature, and a more general Dicke-Heisenberg model.

First, let us consider the Dicke-Ising model, which supplements the Dicke Hamiltonian with nearest-neighbor Ising interactions along the quantization axis,
\begin{align}\label{eq:dickeIsing}
    H_{\rm DI} &= H_{\rm Dicke} - J \sum_{\langle i,j \rangle} \sigma_i^z \sigma_j^z 
    \nonumber \\
    &= H_{\rm Dicke} - 4J \sum_{\langle i,j \rangle} s_i^z s_j^z .
\end{align}
Here we take $\langle i,j\rangle$ as nearest neighbors on a one-dimensional chain with periodic boundary conditions. The global minus sign in the Ising term implies that $J>0$ favors ferromagnetic alignment along the $z$ axis (and $J<0$ antiferromagnetic). We adopt the $zz$ convention~\cite{Rohn2020,zhang2014quantum}, for which the interaction term is diagonal in the computational $S^z$ basis (in this sense ``classical-like''), while the light-matter coupling is transverse (non-commuting), providing a transparent separation between the interaction channel and the Dicke coupling axis. Variants of the Dicke-Ising model with interactions in $x$~\cite{deBernardis2018cavity,cortese2017polariton,sur2026amplified} or $y$~\cite{lee2004first, gammelmark2011phase, gammelmark2012interacting} have also been studied in distinct settings. In the limit $J\to 0$ one recovers the standard Dicke model, while for $g\to 0$ the Hamiltonian reduces to an Ising ring in a longitudinal field.

To probe more complex and anisotropic exchange interactions beyond the Ising limit, more general spin-spin interactions can be considered. Such interactions can be engineered, for example, in Rydberg atom arrays~\cite{Whitlock2017,browaeys2016exp}, circuit QED architectures~\cite{pashkin2003quantum}, and in optical cavities~\cite{davis2020protecting}. The class of extended Dicke models reads
\begin{align}\label{eq:extendedDicke}
H_{\rm Dicke} - \sum_{i<j} \left[ J^x_{i j} s_i^x s_j^x + J^y_{i j} s_i^y s_j^y + J^z_{i j} s_i^z s_j^z \right].
\end{align}
Unlike the Ising case, non-diagonal terms in the computational $S^z$ basis are introduced, which introduces additional quantum channels in the spin sector and provides a natural setting to assess the interplay between direct exchange interactions and the transverse Dicke coupling. Notably, subsets of this model has been recently studied in the context of quantum battery~\cite{dou2022xxz} and of Rydberg arrays~\cite{han2024interaction}.

\subsection{Spin-Holstein models}
Moving beyond collective light-matter interactions, the regime where individual spins couple to distinct, local bosonic modes is also captured by our presented method. The spin-Holstein model is an analog to the Holstein model of condensed matter physics describing electron-phonon interactions. In the context of trapped ions, such models are realized by coupling the internal states of the ions to their local vibrational degrees of freedom, for instance, in arrays of individual microtraps~\cite{knorzer2022spin}. As described in Ref.~\onlinecite{porras2004effective}, applying state-dependent forces to the ions displaces the vibrational modes conditional on the spin configuration, facilitating the simulation of effective spin-spin interactions mediated by the phonons, and allowing for the exploration of polaron physics where the spin and bosonic sectors become entangled.
The paradigmatic Hamiltonian for this system is given by
\begin{equation}\label{eq:sH}
H_{\rm SH} = H_{\rm spin} + \sum_{j=1}^N \omega_j a_j^\dagger a_j + \sum_{j=1}^N g_j s_j^z (a_j + a_j^\dagger ).
\end{equation}
The specific choice of the coupling operator, i.e., transverse~\cite{min2025mixed} or longitudinal~\cite{knorzer2022spin,sun2025quantum} (shown above), depends on the laser configuration~\cite{porras2004effective}. Still, a longitudinal ($\sim S^z x$) coupling is often employed to map the system onto the standard Holstein model~\cite{Holstein1959_1,Holstein1959_2}, where the spin operator mimics the local electron density.

\section{Hybrid numerical approach}
\label{sec:method}

Spin-boson Hamiltonians such as Eq.~\eqref{eq:fullH} are challenging not only because the bosonic sector is infinite-dimensional, but also because the spin and boson degrees of freedom impose qualitatively different numerical demands. Interacting spins may develop strong many-body correlations that typically require exact many-body solvers (e.g., ED or DMRG), which are already limited by memory and entanglement growth. Bosons, in contrast, can often be described efficiently by a simple variational state due to their comparatively weak internal correlations.
However, treating them with a single computational approach typically enforces a hard compromise: one either truncates the bosonic occupation basis or relaxes the spin sector to a variational (or mean-field) description. Both routes can introduce biases which may distort or even eliminate important spin-boson correlations that the coupled problem is meant to capture.

To address these challenges, we represent the ground state in a non-Gaussian variational manifold, which includes a Gaussian bosonic state along with a dressing transformation that captures spin-boson correlations beyond mean-field. This family of variational ans\"atze, called the non-Gaussian states (NGS) and described in detail in~Refs.~\onlinecite{shi2018variational} and \onlinecite{wang2020zero}, reads
\begin{equation}
    \ket{\psi_{\rm NGS}} = U_{\lambda} ( U_{\mathrm{GS}} \ket{0_{\rm b}}\otimes\ket{\phi_{\rm s}} ),
\end{equation}
Here, $U_{\mathrm{GS}}$ is a Gaussian transformation that combines displacement and squeezing, $\ket{0_{\rm b}}$ is the bosonic vacuum and the dressing transformation $U_{\lambda}$ includes correlations between the two distinct degrees of freedom. Unlike purely variational frameworks, here $\ket{\phi_{\rm s}}$ is not a fermionic Gaussian state, but the exact ground state of an effective spin Hamiltonian obtained by averaging out $H$ in $U_{\lambda} U_{\mathrm{GS}} \ket{0_{\rm b}}$~\cite{wang2020zero}. To find the approximate ground state, we optimize the variational parameters and solve the effective spin problem in self-consistent updates until convergence. 

In the following, we discuss each variational layer of the ansatz, namely bosonic Gaussian states (Sec.~\ref{sec:GS}), the dressing transformation (Sec.~\ref{sec:dressing}), and the resulting self-consistent optimization loop in Sec.~\ref{sec:SFopt}.

\subsection{Bosonic Gaussian Layer}\label{sec:GS}

The Gaussian representation provides a robust framework for examining bosonic states in systems where interactions are accurately approximated by quadratic Hamiltonians, which is usually the case at weak coupling. For instance, the thermodynamic-limit zero-temperature ground state of the Dicke model is well described by a mean-field (separable) coherent state~\cite{Kirton2019}.
Coherent states are often regarded as the most classical pure states of the cavity field, since normally ordered operators act on them as if the field amplitude were a complex number. In the limit of large photon numbers they closely reproduce the dynamics of a classical electromagnetic wave, so the field can be approximated by classical wave equations while still retaining quantum features such as vacuum fluctuations.
The Gaussian state is a straightforward generalization that includes squeezed fluctuations.
Mathematically, we take $\ket{\psi_{\rm b}}= U_{\rm GS} \ket{0_{\rm b}}$, where $\ket{0_{\rm b}}$ is the vacuum and
\begin{equation}
    U_{GS} = U_D U_S =  \exp(i R^T \sigma \Delta_R)\exp(-\frac{i}{2} R^T \xi R),
\end{equation}
where $R = (x_1, x_2, \dots, x_M, p_1, p_2, \dots, p_M)^T \equiv (x,p)^T$ is the quadrature vector. This transformation comprises two unitary operators: the displacement operator $U_D$ and the squeezing operator $U_S$, where 
\begin{equation}
    \Delta_{R}
    = (\Delta_{x_1}, \dots, \Delta_{x_M}, \Delta_{p_1},  \dots, \Delta_{p_M} )^T,
\end{equation}
is the $2M$-element displacement vector encoding $\Delta_{x_m} = \langle x_{m} \rangle_{\rm GS}$ and $\Delta_{p_m} = \langle p_{m}\rangle_{\rm GS}$, and
$\xi$ is a real symmetric $2M\times2M$ matrix that parametrizes squeezing, with $\langle \vdot \rangle_{\rm GS} = \bra{\psi_{\rm b}} \vdot \ket{\psi_{\rm b}}$. While the first moments are encoded in $\Delta_{R}$, the second moments are encoded in the covariance matrix
\begin{equation}
    \Gamma = \frac{1}{2} \langle \{ \delta R , \delta R^T \} \rangle_{\rm GS}
    = \frac{1}{2} S S^T,
\end{equation}
where $\delta R = R - \Delta_R$ and $S \in {\rm Sp}(2M)$ is the symplectic matrix defined below.
Under $U_{\mathrm{GS}}$, the relevant operators transform as
\begin{align}
    U_{\mathrm{GS}}^{\dagger}   R   U_{\mathrm{GS}}
    &= S R + \Delta_{R},
    \\
    U_{\mathrm{GS}}^{\dagger}   s_{i}^{\alpha}   U_{\mathrm{GS}}
    &= s_{i}^{\alpha} .
\end{align}
The symplectic matrix $S=e^{\sigma\xi}$, where $\sigma$ is the standard symplectic form, encodes the quadrature squeezing, such that the covariance matrix $\Gamma$ elements can be used as variational parameters directly instead of $\xi$ when appropriate.

In practice, expectation values in Gaussian states can be efficiently computed as they obey Wick’s theorem~\cite{wick1950the}, which allows one to reexpress expectation values of an arbitrary product of mode operators in terms of products of pairs.

\subsection{Dressing transformation and spin-boson entanglement}\label{sec:dressing}

While the Gaussian ansatz described in Sec.~\ref{sec:GS} efficiently parametrizes bosonic fluctuations, it inherently enforces a factorizable structure between the spin and boson sectors. Such product states fail to capture spin-boson entanglement that characterizes the ground state of strongly coupled light-matter systems~\cite{shi2018variational}. To address this, a standard strategy in electron-phonon problems is to map the complexity from the wavefunction onto the Hamiltonian itself via a polaron or Lang-Firsov transformation~\cite{Lang1963}.

In the context of spin-boson models, one can construct an analogous transformation to absorb the linear coupling into the frame definition. Below, we illustrate this strategy using the Dicke model. The canonical transformation is given by a spin-dependent displacement $U = \exp\left(i g' S^{x} p/\omega \right)$. Applying this transformation to \eqref{eq:dicke} yields:
\begin{align}\label{eq:dicke_dressed_H}
    \Tilde{H} = \frac{\omega}{2} (x^2 + p^2) 
    + \varepsilon \left( S^z \cos{\zeta} - S^y \sin{\zeta} \right) - \frac{4g^2}{N\omega} (S^x)^2 ,
\end{align}
where $\zeta = g' p/\omega$.
This transformed Hamiltonian reveals the underlying physics of the strong-coupling regime: the explicit linear spin-photon coupling is eliminated, replaced by photon-mediated all-to-all spin interactions $\propto (S^x)^2$. This structure is particularly useful for various reasons. For example, in this frame, since $a \to a - (g'/\omega) S^x / \sqrt{2}$, the number of bosonic excitations in the ground state is comparatively small, allowing for smaller Hilbert space cutoff with respect to the (bare) lab-frame Hamiltonian~\cite{chen2008numerically,garwola2025compact}. Additionally, direct spin-photon couplings are embedded into higher order $p$-dependent terms, which can be perturbativelly small and are expected to vanish in the thermodynamic limit as $\langle p \rangle \to 0$, making it also useful for deriving effective theories (see Ref.~\onlinecite{romanroche2022effective} for a comprehensive study). Finally, this assumption reproduces the perturbative Schrieffer-Wolff transformation result in the dispersive and fast cavity regimes~\cite{brune1996observing,pilar2020thermodynamics} (see also alternative slow-cavity versions~\cite{jarrett2007optical,jarrett2007quantum}).

This correspondence confirms that a specific frame change provides an improvement towards non-perturbative analysis. However, while such fixed canonical transformation is optimal in the aforementioned limits, it is not universally valid. To capture the physics of regimes where neither the pure mean-field nor the above description is sufficient, we promote the canonical frame change to a unitary transformation embedded in the variational manifold, including $\lambda$ as a variational parameter that controls the spin-boson correlations~\cite{silbey1984variational_1,silbey1984variational_2,Hohenadler2004quantum}. The variational unitary is defined as:
\begin{align}\label{eq:singlemodeU}
    U_{\lambda} &= \exp\left( -\frac{2g}{\omega\sqrt{N}} \lambda (a^\dagger - a) \sum_j s_j^x \right)
    \equiv \exp\left(i \frac{g'}{\omega} \lambda S^{x} p \right).
\end{align}
This allows the solver to interpolate between weakly- ($\lambda \to 0$) and strongly-correlated ($\lambda \to 1$) limits. On one hand, $\lambda=0$ represents a naive separable state ansatz, commonly applied to quantum optics studies where light is treated as a semi-classical coherent field. On the other hand, $\lambda=1$ relates to more a sophisticated approach to eliminate the cavity field in the thermodynamic limit, as explained above. Therefore, promoting the frame change to a variational layer should allow one to better capture spin-boson correlations.

In general, this framework may be extended to a multi-mode, spin-dependent transformation compatible with the general Hamiltonian in Eq.~\eqref{eq:fullH},
\begin{equation}\label{eq:Ulambda_general}
    U_{\lambda}
        = \exp \left(i\sum_{i=1}^{N} \sum_{m=1}^{M} s_{i}^{\mu} \lambda_{i m} p_{m} \right) . 
\end{equation}

Throughout this work, however, we focus on the homogeneous single-mode form relevant to the cavity-QED limit explored in Ref.~\onlinecite{mendonca2025role}, which is given by Eq.~\eqref{eq:singlemodeU}. The non-Gaussian effective Hamiltonian and variational energy functional in the general case are derived and presented in Appendix~\ref{sec:appx_general}.

\subsection{Self-Consistent Optimization}\label{sec:SFopt}

\begin{figure}[t]
    \centering
    \includegraphics[width=\columnwidth]{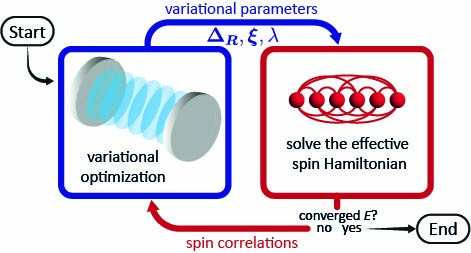}
     \caption{Self-consistent optimization loop. For fixed bosonic variational parameters, the effective spin Hamiltonian is solved (DMRG), yielding spin expectation values and correlations that update the variational energy functional. The procedure is iterated until convergence of the energy and selected observables.}
    \label{fig:method}
\end{figure}

The optimization loop is based on a repeated two-step protocol. For fixed $\{ \Delta_R,\Gamma,\lambda \}$, the variational problem reduces to solving the ground state of the effective spin Hamiltonian $H_{\rm eff}$, which is obtained by averaging out the full Hamiltonian over $U_{\lambda} U_{\mathrm{GS}} \ket{0_{\rm b}} \equiv U_{\lambda} \ket{\psi_{\rm b}}$: 
\begin{align}\label{eq:Heff}
    H_{\rm eff} &= \bra{\psi_{\rm b}} U_\lambda^\dagger H U_\lambda \ket{\psi_{\rm b}}
    \\
    &= \sum_{i=1}^{N}\sum_{\alpha\in\{x,y,z\}}
    h_{i\alpha}^{\rm eff}   s_{i}^{\alpha}
    \nonumber\\
    &+
    \sum_{i=1}^{N}\sum_{j=1}^{N}\sum_{\alpha,\beta\in\{x,y,z\}}
    J_{i j,\alpha \beta}^{\rm eff}   s_{i}^{\alpha}   s_{j}^{\beta},
\end{align}
with coefficients explicitly given in Appendix~\ref{sec:appx_general}. The variational parameters $\{ \Delta_R,\Gamma,\lambda \}$ are, therefore, updated for fixed $\ket{\phi_{\rm s}}$ via minimization of the variational energy functional,
\begin{equation}
    \mathcal{E}(\Delta_R,\Gamma,\lambda) = \langle H \rangle_{\rm NGS} ,
\end{equation}
at iteration step $i$, so variational parameters and spin MPS must be determined self-consistently. This two-step iterative procedure is repeated until convergence~\cite{fehske1995hole,wang2020zero,wang2021phonon,wang2021fluctuating}. The workflow diagram is shown in Fig.~\ref{fig:method}. 

We solve the effective spin problem using DMRG. A practical advantage over direct MPS treatment of the full spin-boson Hamiltonian is that the MPS bond dimension required for the effective spin Hamiltonian is typically much smaller than in a direct simulation, since the bosonic degrees of freedom are incorporated through the variational layer. This is discussed in more detail in Sec.~\ref{sec:benchmark}. While the variational principle imposes the true ground state as the lower bound, we use the Gaussian limit (GS-DMRG, $\lambda=0$) as an upper bound, given the manifold hierarchy $\mathcal{M}_{\rm GS} \subseteq \mathcal{M}_{\rm NGS}$.
This immediately implies energy ordering upon variational minimization, i.e., the best energy attainable within the Gaussian manifold cannot be lower than the one attainable within the non-Gaussian manifold, i.e. $E_{\rm min}^{\rm GS} \ge E_{\rm min}^{\rm NGS} \ge E_0$, where $E_0$ is the true ground state energy. This nesting provides a controlled internal benchmark. Appendix~\ref{sec:appx_manifold} outlines a proposed systematic manifold hierarchy for further accuracy improvements. The GS description also acts as a useful and intuitive approximation, as it neglects correlations between the subsystems.

\section{Results}\label{sec:results}

In this section we benchmark the approach introduced in Sec.~\ref{sec:method} in a representative cQED setting of a single cavity mode coupled to a one-dimensional spin chain with periodic boundary conditions: Dicke and Dicke-Ising models. We show qualitative comparison between NGS-DMRG and reference DMRG, including the Gaussian-limit as a baseline.
Here, \textit{reference DMRG} denotes DMRG performed on the composite spin-photon chain with an explicit photon cutoff $n_{\max}$. We verify convergence by increasing $n_{\max}$ until the ground-state energy changes by less than our target tolerance. We find that a comparatively large cutoff (with respect to $N$) is required to reach this converged regime, reflecting the rapid growth of bosonic occupation at strong coupling. Still, as the spin-photon entanglement decays in both small- and large-$g$ limits, reference DMRG maintains a high degree of efficiency, not requiring LBO or polaron-frame calculations in this specific case.

\subsection{Dicke model}

As a first validation step, in Secs.~\ref{sec:GSsol}--\ref{sec:singlemode} we study the Dicke model given by Eq.~\eqref{eq:dicke}, which acts as controlled baseline to check NGS-DMRG solutions against well established numerical and analytical solutions, and to efficiently quantify finite-size effects across varying number of atoms. Written in terms of collective spins only, due to permutation symmetry, the spin sector reduces to $N+1$ states (instead of $2^N$), which simplifies the problem significantly. This symmetry also implies low entanglement between spins, enabling efficient numerical analysis, although a photon-number truncation is still required for reference DMRG.

\begin{figure}[!t]
    \centering
    \includegraphics[width=0.95\columnwidth]{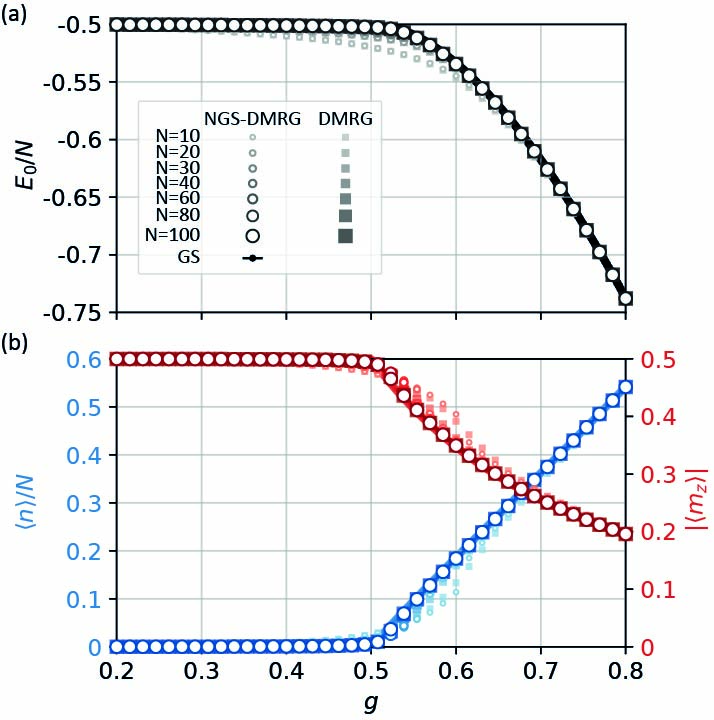}
    \caption{Dicke model observables for the three solvers and various $N$. (a) Average energy, (b) photon number per spin and absolute magnetization as a function of $g$. Marker sizes scale proportionally with the $N$. For reference DMRG, we fix $n_{\max} =$ 40, 40, 60, 60, 80, 100, and 140 for $N =$ 10, 20, 30, 40, 60, 80, and 100, respectively.}
    \label{fig:dicke}
\end{figure}

\subsubsection{Gaussian solution}
\label{sec:GSsol}

As a baseline, we first consider the separable Gaussian limit in which the bosonic mode is described by a displaced vacuum. In practice, this corresponds to a cavity mean-field treatment where the bosonic operator is replaced by a c-number,
$a\to\langle a\rangle\equiv\alpha$ or equivalently $x\to\langle x\rangle\equiv\Delta_x$, and we set $\langle p\rangle=0$ as it minimizes to zero. The resulting effective spin Hamiltonian reads
\begin{equation}\label{eq:dicke_heff_GS}
    H_{\rm eff}(\Delta_x)=\frac{\omega}{2}\Delta_x^2+\varepsilon S^z+g'\Delta_x S^x ,
\end{equation}
where the spin-boson coupling is replaced by a semi-classical (self-consistent) transverse field.
Within the variational framework of Sec.~\ref{sec:method}, this is precisely the $\lambda=\xi=0$ limit, providing a controlled reference point for the following non-Gaussian solutions. Throughout this paper, to not confuse the terminologies, we reserve the term ``mean-field'' for strictly separable states, i.e., excluding both spin-boson and many-body spin-spin entanglement. Here, only spin-photon entanglement is neglected, and we label it Gaussian solution (GS). In both GS and NGS cases, solving an effective spin Hamiltonian determines the wavefunction in the spin sector.

Minimization of the variational energy functional with respect to the field displacement is analytic,
\begin{equation}\label{eq:Deltax_GS}
    \Delta_x^*=-\frac{g'}{\omega}\langle S^x\rangle_{\rm s},
\end{equation}
where $\langle \vdot \rangle_{\rm s} = \bra{\phi_{\rm s}} \vdot \ket{\phi_{\rm s}}$ represent bare spin averages. The variational minimization turns into a self-consistent problem: $\Delta_x^*$ depends on the spin expectation values of the state that minimizes $H_{\rm eff}(\Delta_x)$. In this limit, the resulting energy density can be written solely in terms of collective magnetization,
\begin{equation}\label{eq:dicke_E0_GS}
    \mathcal{E}/N=\varepsilon\left( \langle m_z \rangle - \frac{4g^2}{\omega\varepsilon} \langle m_x \rangle^2\right),
\end{equation}
with $m_\alpha = S^\alpha /N$ and $\varepsilon$ acts as the unit of energy. Longitudinal magnetization $\langle m_z \rangle$ survives until $4g^2/(\omega\varepsilon)=1$ where the system undergoes the well-known second-order phase transition to the superradiant state where $\langle n \rangle/N \propto \langle m_x \rangle^2 > 0$. Therefore, we identify $g_c=\sqrt{\omega\varepsilon}/2$ as the phase transition point, which is numerically confirmed in Fig.~\ref{fig:dicke} through the self-consistent Gaussian solution.  This exactly matches the mean-field theory solution shown in Appendix~\ref{sec:appx_MFT} and the literature~\cite{reslen2005direct,vidal2006finite}, confirming its validity in the thermodynamic limit. However, more general models featuring spin-spin interactions are not expected to follow this behavior.

Finally, although Eq.~\eqref{eq:dicke_heff_GS} was motivated through a coherent-state (displacement-only) picture, allowing for a fully Gaussian bosonic state leads to the same result, since the squeezing parameter analytically minimizes to $\xi=0$. As we show next, squeezing becomes relevant in combination with non-Gaussian dressing, particularly at finite sizes near the transition.

\subsubsection{NGS Solutions}
\label{sec:singlemode}

Applying the NGS framework to the homogeneous single-mode Dicke interaction allows us to first simplify the non-Gaussian manifold using physically motivated constraints. Because the coupling is uniform, we represent the dressing transformation as a collective spin rotation about the $x$ axis, $\lambda_{j1}^{\alpha}=\lambda (g'/\omega) \delta_{\alpha x}$. In the bosonic sector, the restriction to a single mode naturally reduces the displacement vector and covariance matrix to two elements and a $2\times2$ matrix, respectively. We can further simplify this by noting that the momentum quadrature $p$ appears only quadratically in the uncoupled cavity Hamiltonian; thus, any finite momentum displacement strictly increases the energy, allowing us to safely set $\langle p\rangle\equiv \Delta_p=0$. Additionally, assuming the bosonic fluctuations are dominated by squeezing along the principal axes (neglecting quadrature rotations), the covariance matrix becomes strictly diagonal, $\Gamma=\mathrm{diag}\{e^{2\xi}/2,e^{-2\xi}/2\}$, parameterized entirely by a single scalar $\xi$. Finally, because the position displacement $\Delta_x$ appears at most quadratically in Dicke-type models, it can be minimized analytically ($\partial \mathcal{E}/ \partial \Delta_x = 0$) and eliminated from the problem entirely, yielding
\begin{equation}\label{eq:Deltax}
    \Delta_x^* = -\frac{(1-\lambda)g'}{\omega} \langle S^x \rangle_{\rm s}.
\end{equation}
We are thus left with a low-dimensional variational problem in a self-consistent loop: solve the effective spin Hamiltonian
\begin{align}\label{eq:dickeHeff_NGS}
    H_{\rm eff} = \varepsilon \Bigg[ &\exp\left( - \frac{2\lambda^2 g^2}{N\omega^2} e^{-2\xi} \right) S^z - \frac{8g^2}{N\omega\varepsilon} (1 - \lambda)^2 \langle S^x \rangle_{\rm s} S^x 
    \nonumber\\ &+ \frac{4g^2}{N\omega\varepsilon} (\lambda^2 - 2\lambda) (S^x)^2 \Bigg] ,
\end{align}
and numerically minimize the corresponding variational energy functional, 
\begin{align}
    \mathcal{E} &= \omega \sinh^2\xi + \varepsilon \exp\left( - \frac{2\lambda^2 g^2}{N\omega^2} e^{-2\xi} \right) \langle S^z \rangle_{\rm s} 
    \nonumber\\ &- \frac{4g^2}{N\omega} (1 - \lambda)^2 \langle S^x \rangle_{\rm s}^2 + \frac{4g^2}{N\omega} (\lambda^2 - 2\lambda) \langle (S^x)^2 \rangle_{\rm s} ,
\end{align}
with respect to the two parameters $\{\xi,\lambda\}$. Since only two variational parameters are involved, we do not require elaborate protocols and employ standard numerical optimization routines. The homogeneous single-mode restriction defines the minimal non-Gaussian manifold considered in this paper, recently shown to be very accurate for extended Dicke models~\cite{mendonca2025role}.

Comparing effective Hamiltonian~\eqref{eq:dickeHeff_NGS} to the Gaussian baseline in Eqs.~\eqref{eq:dicke_heff_GS} and \eqref{eq:dicke_E0_GS} highlights the physical mechanism behind the non-Gaussian improvement. While the Gaussian ansatz only captures a self-consistent mean-field shift---acting as an effective transverse field proportional to $\langle S^x \rangle_{\rm s} S^x$---the dressing transformation explicitly generates a two-body, photon-mediated, infinite-range spin term proportional to $(S^x)^2$. This arises because all spins are coupled to the global cavity field. The inclusion of these photon-mediated processes generates spin-spin correlations absent in both Gaussian and mean-field approaches. The resulting expression for the photon number is also strongly affected, as  $\langle n \rangle_{\rm NGS} \propto a \langle S^x \rangle_{\rm s}^2 + b \langle (S^x)^2 \rangle_{\rm s}$, while $\langle n \rangle_{\rm GS} \propto \langle S^x \rangle_{\rm s}^2$.

Figure~\ref{fig:dicke} shows ground-state observables for multiple number of atoms as a function of the light-matter coupling $g$. The non-Gaussian energies in Fig.~\ref{fig:dicke}(b) lie below the Gaussian baseline for all finite sizes, reflecting the expected variational improvement. As $N$ increases, the energetic gain decreases and the two manifolds approach each other, consistent with the fact that the separable cavity mean-field description becomes increasingly accurate in the large-$N$ limit. At large $g$ the two approaches also become nearly indistinguishable, as we enter the semi-classical limit.

The order parameters in Fig.~\ref{fig:dicke}(c) follow the expected trend: for small $g$ the photon number is suppressed while the magnetization is close to its maximal value; beyond the crossover/critical region the photon number grows and the longitudinal magnetization is reduced by the cavity-induced transverse field. Both observables exhibit the expected finite-size smoothing of the transition.

As an additional numerical validation, we compare against full DMRG simulations of the original Dicke Hamiltonian, where the bosonic Hilbert space is made finite by truncating the photon number to $n_{\max}$. For the small sizes shown, we find close agreement once $n_{\max}$ is chosen large enough to ensure convergence.

Although the general formalism allows for mode- and site-dependent variational parameters, the homogeneous single-mode restriction already captures the essential Dicke-type spin-photon correlations while keeping the optimization remarkably simple. In the next section, we show that the same setting yields accurate results for interacting extensions such as the Dicke-Ising chain.

\subsection{Dicke-Ising model}

\begin{figure}[t]
    \centering
    \includegraphics[width=\columnwidth]{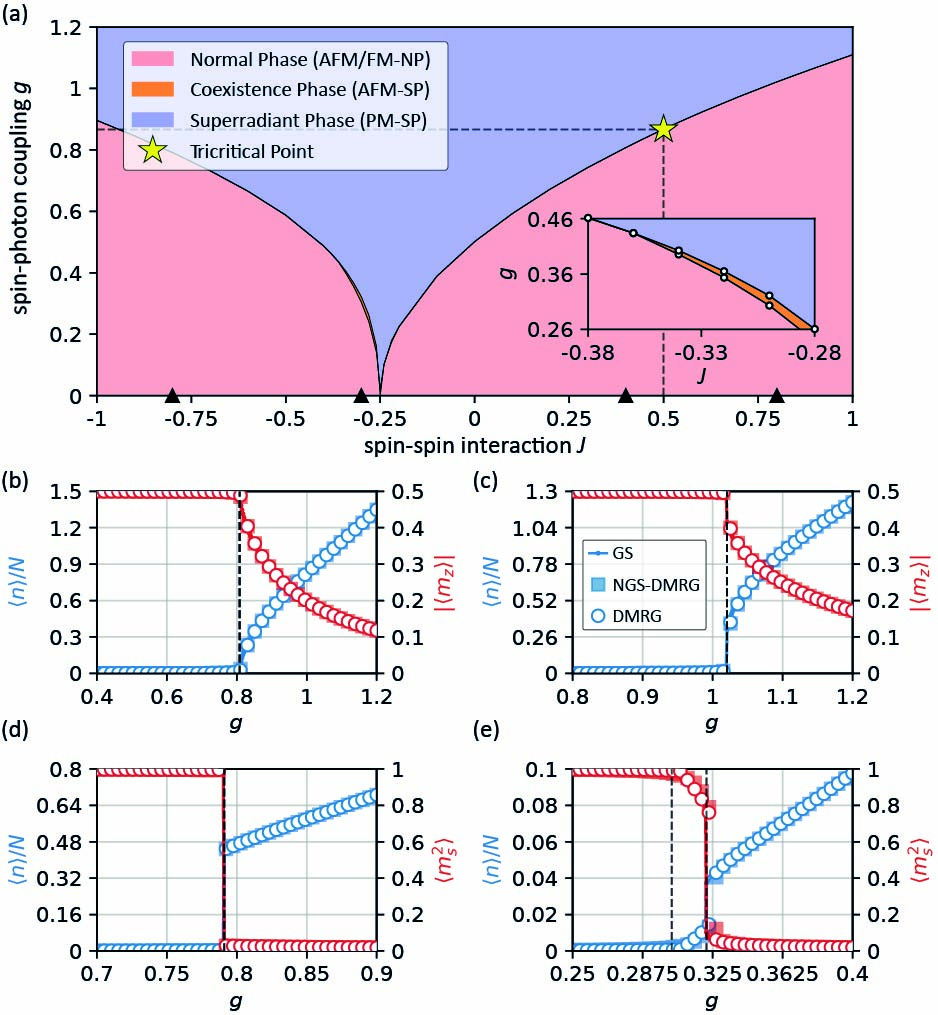}
    \caption{Dicke-Ising model phase diagram and observables. (a) Phase diagram in the Gaussian limit as a function of the spin-spin interaction $J$ and spin-photon coupling $g$, depicting first- and second-order phase boundaries alongside a tricritical point. The mapped regions include the normal phases (AFM/FM-NP), the superradiant phase (PM-SP), and a coexistence phase (AFM-SP) detailed in the inset. (b)-(e) System observables evaluated along vertical cuts at fixed $N=100$ for representative $J$ values, indicated by triangles in (a). The panels correspond to (b) $J=0.4$, (c) $J=0.8$, (d) $J=-0.8$, and (e) $J=-0.3$. Results from the three solvers are shown for comparison in each panel. Reference DMRG data is obtained with a maximum photon number $n_{\max}=350$.}
    \label{fig:DI_diagram}
\end{figure}

Having established our method in the previous section, we now apply it to a physical scenario where competing interactions play a decisive role. The model examined below extends the standard Dicke Hamiltonian by introducing nearest-neighbor terms that compete or cooperate with the spin-photon interactions. In this regime, the cavity field behaves non-trivially, in principle demanding a method capable of capturing correlations beyond the mean-field level. While the presented method would allow for a broader range of spin interactions, such as long- and infinite-range couplings~\cite{han2024interaction,koziol2025melting}, and more complex extended Dicke models~\cite{peraca2024quantum}, the Dicke-Ising model described by Eq.~\eqref{eq:dickeIsing} already probes a broad range of regimes where short-range exchange competes with cavity-induced collective interactions. We assess our results by comparison to reference DMRG and GS ansatz treatments. 

The Ising interaction allows us to tune between ferromagnetic (FM) and antiferromagnetic (AFM) microscopic interactions by varying $J$. In the absence of coupling to light (for $g=0$), the transition between the two phases occurs at $J_c = -\varepsilon/4$. Accordingly, we organize the analysis into two regimes: the ferromagnetic regime ($J>J_c$) and the antiferromagnetic regime ($J<J_c$).

The interplay between the atomic splitting $\varepsilon$ (set as the energy unit), spin-photon coupling $g$, and the Ising interaction strength $J$ generates a rich phase diagram governed by distinct transition behaviors~\cite{langheld2025quantum}. We establish a rigorous benchmarking protocol by extrapolating to the thermodynamic limit, which is identified as the Gaussian baseline as spin-photon correlations become negligible. The presented analysis reveals the full complexity of the global parameter space, allowing us to map out the phase diagram featuring both first- and second-order phase transitions, and a variety of phases, as shown in Fig.~\ref{fig:DI_diagram}(a).

\subsubsection{FM regime}\label{sec:dicke_ising_FM}

We start in the ferromagnetic regime $J>J_c$, where the Ising coupling favors spin alignment along $z$. 
For moderate values of $J\lesssim 0.5$, the normal state remains adiabatically connected to the Dicke normal phase and the main effect of the Ising coupling is the renormalization of the atomic excitation gap as explained below. 
This provides a simple analytic benchmark for the numerical algorithm.

To estimate the shifted critical coupling, we decouple the Ising interaction around the fully polarized normal state,
\begin{equation}
    \sigma_i^z \sigma_j^z \approx 
    \sigma_i^z \langle \sigma_j^z \rangle 
    + \langle \sigma_i^z \rangle \sigma_j^z 
    - \langle \sigma_i^z \rangle \langle \sigma_j^z \rangle,
\end{equation}
and use $\langle\sigma^z\rangle\simeq -1$ in the normal phase, which yields
\begin{equation}
    \sigma_i^z \sigma_j^z \approx -2\sigma_i^z - 1.
\end{equation}
For a 1D chain with periodic boundary conditions and using our normalization of the Ising term, the spin Hamiltonian becomes, up to an additive constant,
\begin{equation}
    \frac{\varepsilon}{2}\sum_i\sigma_i^z 
    - J\sum_{\langle i,j\rangle}\sigma_i^z\sigma_j^z
    \approx
    \left(\frac{\varepsilon}{2}+2J\right)\sum_i\sigma_i^z,
\end{equation}
and is fully characterized by an effective atomic splitting $\varepsilon_{\rm eff}=\varepsilon+4J$. 
Applying the standard Holstein-Primakoff analysis with $\varepsilon\to\varepsilon_{\rm eff}$ we then predict a shifted threshold (see also Appendix~\ref{sec:appx_MFT})
\begin{equation}\label{eq:gcJ}
    g_c^2(J)=\frac{\omega \varepsilon_{\rm eff}}{4} = \frac{\omega\varepsilon}{4} + \omega J.
\end{equation}
In this regime, Eq.~\eqref{eq:gcJ} provides a direct reference for validating both the GS solution and its NGS improvement for finite-size effects. Figure~\ref{fig:DI_diagram}(b) confirms that the transition remains continuous and shows the dominant effect as the expected shift of the critical coupling, as shown via $\langle n \rangle/N$ and $\langle m_z \rangle$. 

For sufficiently strong ferromagnetic Ising coupling, the transition from the normal phase to the superradiant phase becomes discontinuous, signaling a tricritical point~\cite{leibig2026quantitative}. Its exact location, $J = \varepsilon/2$, has been analytically derived recently in Ref.~\onlinecite{koziol2026} via effective descriptions---related to Eq.~\eqref{eq:dicke_dressed_H} with $\zeta=0$, leading to modified Eqs.~\eqref{eq:dicke_E0_GS} and \eqref{eq:gcJ}--- matching our numerical prediction. Figure~\ref{fig:DI_diagram}(c) shows the characteristic signatures of a first-order transition in our system, such as a discontinuous jump in the order parameters, $\langle n \rangle$ and $\langle m_z \rangle$, after the critical coupling.

In the limit $\varepsilon=0$, this behavior is expected: an additional spin-flip symmetry implies that two distinct symmetry-broken phases can meet through a first-order boundary. More generally, the coexistence of both first- and second-order quantum phase transitions in distinct Dicke-Ising models has been reported decades ago~\cite{lee2004first,gammelmark2011phase,gammelmark2012interacting}---although the thermodynamics of these variants can be computed exactly--- has been recently predicted in a transverse field setting~\cite{romanroche2025bound}, and also in the class of three-level Dicke models~\cite{padilla2023tri}.

\subsubsection{AFM regime}\label{sec:dicke_ising_AFM}

We now turn to the antiferromagnetic case $J<J_c$, where the competition between antiferromagnetic Ising interactions ($J<0$) and the longitudinal field $\varepsilon$ favors an AFM-ordered state at small $g$, with a vanishing photon number. 
In this regime, a direct continuous connection between antiferromagnetic-normal (AFM-NP) and paramagnetic-superradiant phases (PM-SP) is not expected within Landau theory. Indeed, a direct transition between phases that break distinct $Z_2$ symmetries is expected to be of first order.

To characterize the magnetic order in this regime, we introduce the staggered magnetization $m_s = \frac{1}{N} \sum_i (-1)^i \sigma_i^z$. In finite-size NGS-DMRG calculations, the absence of spontaneous symmetry breaking can make the direct evaluation of the one-point function $\langle m_s \rangle$ impractical as an order parameter. A more robust approach relies on two-point correlation functions. We therefore compute the expectation value of the squared operator, which reads
\begin{align}
    \langle m_s^2 \rangle &= \frac{1}{N^2} \sum_{i,j} (-1)^{i-j} \langle \sigma_i^z \sigma_j^z \rangle
    \\
    &= \left(\frac{2}{N}\right)^2 \sum_{i,j} (-1)^{i-j} \langle s_i^z s_j^z \rangle,
\end{align}
where $s_i^z = \sigma_i^z/2$ are the spin-$1/2$ operators. With this normalization, $\langle m_s^2 \rangle$ is of order $\mathcal{O}(1)$ in an ordered AFM phase and scales as $\sim 1/N$ in an uncorrelated state. This observable is strictly related to the longitudinal spin structure factor, corresponding to its value at the antiferromagnetic wavevector, $q=\pi$. 

Figure~\ref{fig:DI_diagram}(d) shows that AFM order and the normal phase remain locked together throughout the small-$g$ region, i.e., $\langle m_s^2 \rangle = 1$ and and the photon number vanishes.
At the critical coupling, both observables change abruptly, as $\langle m_s^2 \rangle$ collapses to $\sim 1/N$ and the photon number becomes a finite growing value, signaling a discontinuous transition into the paramagnetic-superradiant phase.

A separate subtlety in the AFM regime is the appearance of an intermediate phase with both photon condensation and a $z$-ordered spin state. In fact, a phase in which antiferromagnetic order and superradiance coexist, labeled as antiferromagnetic-superradiant (AFM-SP), has been predicted via mean-field theory analysis a decade ago~\cite{zhang2014quantum,gelhausen2016quantum}. While mean-field or perturbative solutions lead to a broad intermediate region, missing the correct phase boundaries, more precise numerical approaches including large scale QMC calculations predict that in one dimension such AFM-SP regime exists in an extremely narrow window~\cite{langheld2025quantum,schellenberger2026infinity}. Rather than a direct first-order boundary between the AFM-NP and PM-SP phases, we observe a distinct parameter window, shown in the zoomed-in phase diagram in the inset of figure~\ref{fig:DI_diagram}(a), where superradiance ($\langle n \rangle/N > 0$) coexists with antiferromagnetic order. In Fig.~\ref{fig:DI_diagram}(e) we show the system undergoes a continuous onset of photon occupation and a finite $\langle m_s^2 \rangle < 1$, signaling a second-order boundary into the coexisting AFM-SP phase, followed by a discontinuous jump where $\langle m_s^2 \rangle \sim 1/N$ as the system enters the PM-SP phase. 

\subsubsection{Summary}

In summary, we applied the NGS-DMRG numerical framework to characterize the phases of the well-known Dicke-Ising model. We observed the entire variety of phases and QPTs reported in recent works. NGS-DMRG overlaps well with reference DMRG for $N=100$, showing that our method is capable of reproducing results of established numerical methods. Additionally, while NGS-DMRG accurately captures finite-size fluctuations, the GS limit leads to intensive observables, even though the bond dimension is not fixed to unity. This is expected, since neglecting spin-photon correlations, i.e., fixing $a\to \alpha$, is compatible with vanishing spin-spin correlations, as $\langle (S^x)^2 \rangle \to 2\langle S^x \rangle S^x$. Moreover, NGS-DMRG and reference DMRG calculations converge towards the GS limit for increasingly large $N$, indicating the connection to the thermodynamic limit, in which the Dicke-Ising model solution features no spin-photon entanglement. The NGS ansatz allows for a better description of fluctuations away from the thermodynamic limit, and includes the possibility of strongly correlated spin-photon states that may develop in certain physical scenarios, which cannot be captured by effective approaches where the bosonic degrees of freedom are eliminated. It was shown in Ref.~\cite{mendonca2025role} that, for example, the Dicke-XXZ model does not converge towards the GS baseline ($\lambda=0$), and therefore, the thermodynamic limit is expected to be connected to the $\lambda=1$ case (with vanishing fluctuations $\zeta=0$)~\cite{romanroche2022effective}.

\section{Quantitative Benchmark}\label{sec:benchmark}

\begin{figure}[t]
    \centering
    \includegraphics[width=0.85\columnwidth]{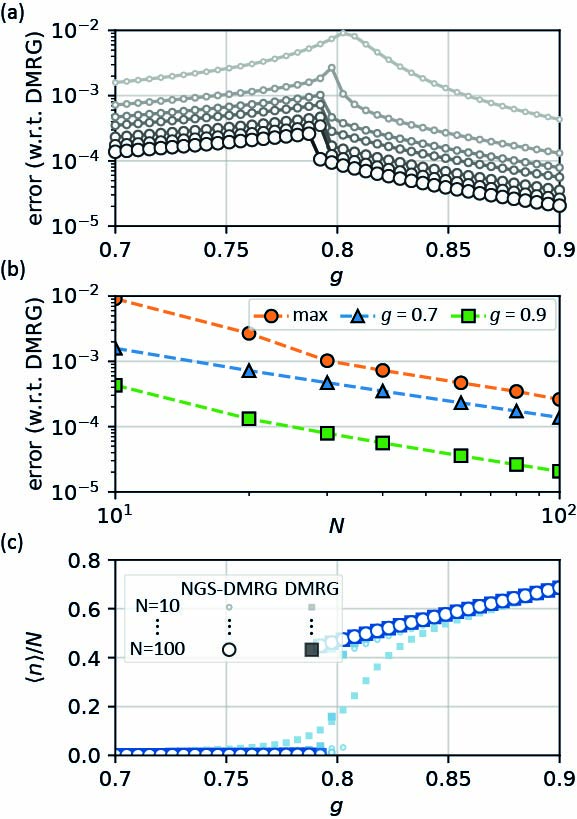}
    \caption{Accuracy of the NGS-DMRG approach for the Dicke-Ising model for $J=-0.8$ and various $N$. (a) Ground-state energy error relative to the reference DMRG solution as a function of the spin-photon coupling $g$. (b) Energy error evaluated at selected coupling strengths as a function of $N$. (c) Photon number per spin $\langle n \rangle/N$ versus $g$ for both NGS-DMRG (circles) and reference DMRG (squares). Marker sizes scale proportionally with the $N$. For the reference DMRG baseline, the photon cutoff $n_{\max}$ is chosen according to the number of atoms as $n_{\max} = 30, 60, 80, 160, 160, 300,$ and $350$ for $N = 10, 20, 30, 40, 60, 80,$ and $100$, respectively.}
    \label{fig:accuracy}
\end{figure}

\begin{figure}[t]
    \centering
    \includegraphics[width=0.85\columnwidth]{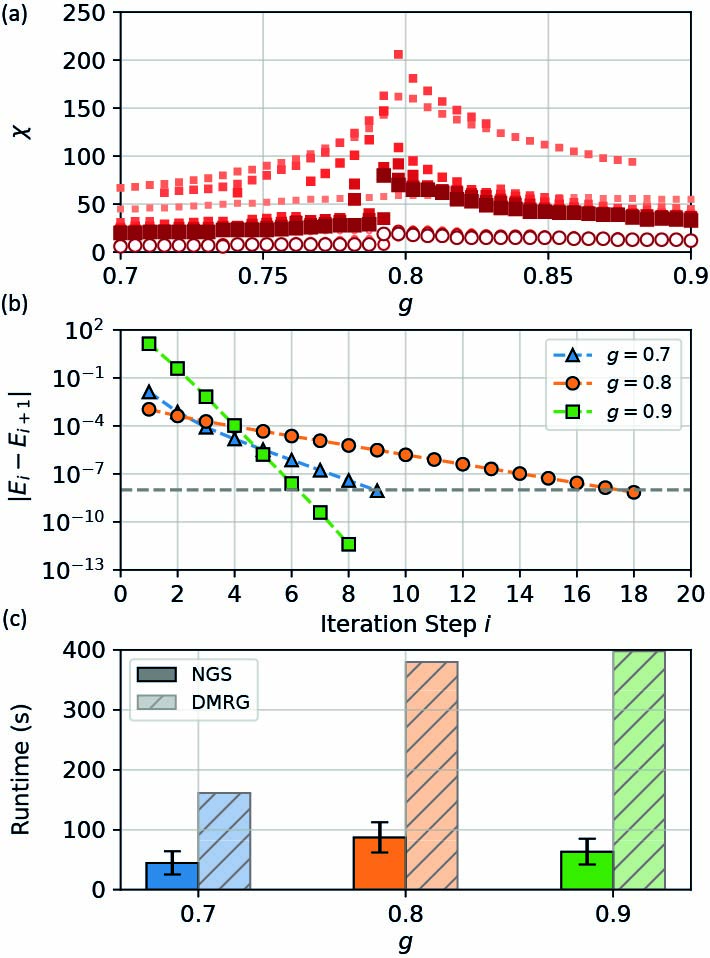}
    \caption{Computational cost comparison for the Dicke-Ising model for $J=-0.8$. (a) Maximum MPS bond dimension $\chi$ as a function of $g$ for the NGS-DMRG (circles) and reference DMRG (squares) across distinct $N$. Marker sizes scale proportionally with $N$. (b) Energy difference per iteration step $i$ demonstrating exponentially fast convergence of the self-consistent NGS loop for $N=100$ and selected values of $g$. (c) Total execution runtime comparing the NGS-DMRG and reference DMRG solvers for $N=100$ and selected values of $g$. For the reference DMRG baseline, the photon cutoff $n_{\max}$ is chosen according to the number of atoms as $n_{\max} = 30, 60, 80, 160, 160, 300,$ and 350 for $N = 10, 20, 30, 40, 60, 80,$ and 100, respectively. 500 realizations were employed to minimize initialization biases in NGS runtime.}
    \label{fig:cost}
\end{figure}

In this section, we compare the accuracy and computational performance of the NGS-DMRG framework against reference DMRG. For the following analysis, we focus on the representative case of $J=-0.8$, zoomed in around the critical coupling $g_c$. While fluctuations and entanglement peak near the phase transition, the total photon number remains sufficiently moderate to ensure that the explicit photon cutoff $n_{\max}$ in the reference DMRG solver remains well controlled and accurate, allowing for a direct comparison.

The two methods diverge fundamentally in their construction of the physical state, leading to distinct numerical challenges. In the reference DMRG approach, the MPS must explicitly represent the composite state by appending a bosonic qudit site of dimension $n_{\max}+1$ to the spin-$1/2$ chain. In the MPS/MPO language, coupling this global bosonic mode to all spins requires long-range connections all across the tensor network, which may result in a growth in the required bond dimension $\chi$ to capture the system's entanglement entropy. Conversely, the NGS-DMRG framework captures spin-boson correlations explicitly through the non-Gaussian unitary transformation $U_\lambda$ prior to the MPS optimization. By variationally tracing out the bosonic degrees of freedom, the tensor network is restricted to an effective spin-only Hamiltonian. However, this theoretical simplification introduces its own numerical bottlenecks. The effective Hamiltonian generates an infinite-range, photon-mediated spin-spin interaction proportional to $(S^x)^2$, and the framework requires an outer self-consistent loop to repeatedly solve this Hamiltonian to optimize the variational parameters. Whether the elimination of the explicit bosonic qudit, and effectively handling spin-boson entanglement variationally, compensates for the algorithmic overhead of repeated infinite-range DMRG calculations is exactly what the following numerical benchmark addresses.

Figure~\ref{fig:accuracy} establishes the quantitative accuracy of the NGS-DMRG framework by comparing its ground-state solutions against the reference baseline. As shown in Fig.~\ref{fig:accuracy}(a), the energy error $|E_0 - E_0^{\rm ref}|$ remains well-bounded across the parameter space, though it peaks in the vicinity of the critical coupling $g_c$, where quantum fluctuations and finite-size effects are most pronounced. Crucially, this discrepancy is not a fundamental limitation of the ansatz in the macroscopic limit. Figure~\ref{fig:accuracy}(b) demonstrates that the energy error decays algebraically with increasing number of atoms $N$ across all analyzed regimes, confirming that the non-Gaussian framework systematically approaches the exact solution. This convergence is further reflected in the physical properties of the system. In Fig.~\ref{fig:accuracy}(c), the normalized photon number $\langle n \rangle/N$ computed via NGS-DMRG shows excellent agreement with the reference solver. While finite-size deviations are visible for the smallest systems near the phase boundary, the order parameters match with high fidelity at sufficiently large $N$. Consequently, the framework successfully captures the correct physical state with only a marginal energy penalty, justifying the transition to the performance and cost analysis.

Having established the physical fidelity of the method, Figure~\ref{fig:cost} addresses the computational trade-offs. Figure~\ref{fig:cost}(a) tracks the maximum MPS bond dimension $\chi$ of the converged state. The NGS-DMRG framework maintains a low and stable bond dimension across the entire parameter range ($g \in [0.7, 0.9]$), remaining well below the reference solver for all analyzed number of atoms. Furthermore, the reference DMRG develops a pronounced cusp around $g_c$, reflecting the severe spin-spin and spin-boson entanglement cost near criticality. However, the reduced tensor network complexity of the NGS-DMRG solver must be weighed against the algorithmic overhead of its self-consistent loop. Figure~\ref{fig:cost}(b) resolves this by tracking the energy difference per iteration step. The roughly linear decrease on a logarithmic scale demonstrates that the outer optimization converges exponentially fast across all parameter regimes. The combination of a suppressed bond dimension and rapid convergence yields a substantial reduction in overall execution time, as confirmed by the runtime comparison in Figure~\ref{fig:cost}(c), although performance of both algorithms could presumably be optimized further~\footnote{To minimize hardware-induced biases in the runtime calculations, simulations were executed with a fixed resource allocation of 16 GB of memory and 24 CPU cores on the Helios supercomputer (PLGrid infrastructure). Both solvers (i) started from a random initial condition, (ii) employed a DMRG cutoff of $10^{-8}$, and (iii) stopped at an energy convergence tolerance of $10^{-8}$ with a minimum of ten DMRG sweeps to guarantee that a well converged MPS is obtained before variational parameter optimizations. Such restriction was employed only to produce this figure.}. 

While this benchmark focuses on the representative case of $J=-0.8$, those gains hold systematically across the phase diagram. As shown in Appendix~\ref{sec:appx_benchmark}, the relative performance disparity scales with the physical complexity of the examined region. In highly complex regimes, such as the intermediate coexistence phase at $J=-0.3$, reference DMRG faces severe scaling and corresponding slow convergence, making the NGS-DMRG advantage even more extreme. Conversely, in less complex parameter regions like the continuous transition at $J=0.4$, the bond dimension requirements for both solvers remain comparatively modest, naturally reducing the performance gap between the two approaches.

\section{Concluding Remarks}\label{sec:conclusion}

We introduced a hybrid variational framework to solve interacting spin-boson Hamiltonians, addressing regimes where conventional simulations are limited by the infinite-dimensional bosonic Hilbert space and strong many-body correlations. The method captures the bosonic state and spin-boson entanglement through a compact non-Gaussian variational manifold beyond mean-field or perturbative approximations. The subsequent spin part of the ground state is the solution of an effective spin Hamiltonian efficiently solved via DMRG. This two-step self-consistent optimization loop avoids the need for an explicit truncation of the bosonic number and retains important spin-boson correlations.
To validate the approach, we selected the Dicke-Ising model as a case study. As a widely studied extension of the standard Dicke Hamiltonian, its rich phase diagram provides a rigorous environment to evaluate the solver across a diverse range of physical regimes and competing interactions. Through this analysis, we successfully mapped a large variety of phases, accurately reproducing the complex boundaries and multicritical behaviors reported in the literature. By capturing this detailed phenomenology, we demonstrated the physical robustness and computational efficiency of our method relative to established numerical algorithms, specifically highlighting a substantial reduction in the required tensor network bond dimension and corresponding runtime. 

We provide an open-source numerical package that not only reproduces the results presented here but already supports a broader class of spin-spin interactions, including arbitrary-range couplings, spatial inhomogeneity, and disorder. Moreover, since the numerical framework is designed to accommodate multi-mode cavity and spin-phonon systems, our next step is to expand the software to include these models. In addition, the modular design of the self-consistent loop makes it straightforward to swap the DMRG backend for alternative solvers, such as exact diagonalization, projected entangled pair states (PEPS)~\cite{Cirac2021}, neural quantum states~\cite{Salmeron2026} or quantum Monte Carlo, which will be implemented in the future to handle higher-dimensional spin geometries. Finally, adapting the method to capture time evolution and dissipation opens the door to exploring non-equilibrium dynamics in strongly correlated open quantum systems.

\section*{Acknowledgments}
K. J. thanks J. A. Koziol and K. P. Schmidt for fruitful discussions regarding the critical behavior of the Dicke-Ising model and its further extensions.
The NGS-DMRG numerical method presented in this work implements DMRG computations using the open-source package \texttt{ITensors.jl} \cite{itensor} as backend. 
K. J. and J. P. M. acknowledge the support by the Narodowe Centrum
Nauki under Grant No. 2023/50/E/ST2/00138.
Y. W. acknowledges the support by the U.S. Department of Energy, Office of Science, Basic Energy Sciences, under Early Career Award No. DE-SC0024524.
The ``Quantum Optical Technologies'' (FENG.02.01-IP.05-0017/23) project is carried out within the Measure 2.1 International Research Agendas programme of the Foundation for Polish Science, co-financed by the European Union under the European Funds for Smart Economy 2021-2027 (FENG). 
We gratefully acknowledge Polish high-performance computing infrastructure PLGrid (HPC Center: ACK Cyfronet AGH) for providing computer facilities and support within computational grant no. PLG/2025/018620.

\section*{Data and Code Availability}
The datasets generated and analyzed during the current study~\cite{mendonca2026dataset_variational}, along with the open-source code package~\cite{mendonca2026software_ngs}, are publicly available.

\appendix

\section{Mean-field Variational Ansatz}\label{sec:appx_MFT}

In the mean-field theory (MFT) approach, the composite light-matter ground state is separable, so as the spin subsystem. Therefore, the photon field is assumed to be coherent while the spin ground state is fully determined by a global rotation on the lowest-lying state. Therefore, the mean-field ansatz reads
\begin{equation}\label{eq:separable_ansatz}
    \ket{\psi} = e^{-i\Delta_x p} \ket{0} \otimes e^{-i\phi S^y} \ket{j,-j},
\end{equation}
where $e^{-i\Delta_x p} \ket{0}$ is a coherent state. Upon minimization, we obtain the same $\Delta_x$ as in Eq.~\eqref{eq:Deltax_GS}, and $\cos{\phi}=(g_c/g)^2$ for $g>g_c$ and $\phi=0$ for $g<g_c$. The two phases separated by $g_c$ can be characterized by spin and photon order parameters. Therefore, 
\begin{equation}
\label{mft_n}
    \frac{\langle n\rangle}{N} =
    \begin{cases}
    0, & g \le g_c,
    \\
    \dfrac{1}{2\omega^2} \left(g^{2}-\dfrac{g_c^{4}}{g^{2}}\right), & g \ge g_c,
    \end{cases}
\end{equation}
showing that the cavity goes from a vacuum state to a photon condensation for $g>g_c$. Furthermore, the spins are aligned until the transition to a paramagnetic state:
\begin{equation}
    m_z =
    \begin{cases}
    -\dfrac{1}{2}, & g \le g_c,
    \vspace{0.25cm}
    \\
    -\dfrac{1}{2} \left(\dfrac{g_c}{g}\right)^{2}, & g \ge g_c.
    \end{cases}
\end{equation}
Additionally, when $g\gg\omega,\varepsilon$ we obtain $\phi=\pi/2$, corresponding to the eigenstate of $S^x$, as expected due to the alignment towards the field. Finally, the minimized energy is given by
\begin{align}
\label{mft_en}
    \frac{E_{\rm min}}{N} =
    \begin{cases}
        -\dfrac{\varepsilon}{2}, & g \le g_c,
        \vspace{0.25cm}
        \\
        -\dfrac{ g^4 + g_c^4 }{\omega g^2}, & g \ge g_c.
    \end{cases}
\end{align}

When the light-matter composite ground state is assumed to be separable, the observables, magnetization and photon number [Eqs.~\eqref{mft_n}-\eqref{mft_en}] do not depend explicitly on $N$. However, in the non-Gaussian approach, the observables may become $N$-dependent due to the inclusion of correlations and fluctuations present away from the thermodynamic limit.

\section{Multi-Mode Generalization}\label{sec:appx_general}

To analyze the structure of the Hamiltonian, we first need to determine how the dressing transformation $U_{\lambda}$ acts on the bosonic and spin operators. Restricted to the single coupling axis $\mu$, this transformation acts as a displacement on bosonic quadratures according to
\begin{equation}
    U_{\lambda}^{\dagger} x_{n} U_{\lambda}
    = x_{n} - \sum_{i=1}^{N} \lambda_{i n} s_{i}^{\mu},
\end{equation}
while $p_{n}$ remains unchanged.
By construction, for each spin at site $i$, a rotation angle
\begin{equation}
    \theta_{i}
    = \sum_{m=1}^{M} \lambda_{i m} p_{m},
\end{equation}
determined by the $p_m$ quadratures, defines a local rotation around the fixed axis $\mu$. We then form a constant antisymmetric matrix,
\begin{equation}
    (\eta^{\mu})^{\alpha\beta}
    =
    \epsilon_{\alpha\beta\mu},
\end{equation}
that serves as the generator of spin rotations, so that the full rotation applied to each spin is
\begin{equation}
    \mathcal{R}_{i}
    =
    \exp(-\theta_{i} \eta^{\mu})
    = \mathbb{1} - \sin{\theta_i} \eta^{\mu}
    + (1 - \cos{\theta_i}) (\eta^{\mu})^{2}.
\end{equation}
This corresponds to the Rodrigues formula but with a flipped angle sign compared to usual definitions.
Therefore, under $U_{\lambda}$ each spin operator $s_{i}^{\alpha}$ transforms as
\begin{equation}
    U_{\lambda}^{\dagger}   s_{i}^{\alpha}   U_{\lambda}
    = \sum_{\beta\in\{x,y,z\}} \mathcal{R}_{i}^{ \alpha\beta}   s_{i}^{\beta}.
\end{equation}

With this in hand we can compute the dressed Hamiltonian. It reads (keeping $\sqrt{2}$ explicit, and extending the bare interaction sum to $\frac{1}{2}\sum_{i,j=1}^{N}$ with $J_{ii}^\gamma \equiv 0$)
\begin{widetext}
\begin{align}
    \widetilde{H} &= 
    U_{\lambda}^{\dagger} H U_{\lambda}
    \nonumber\\
    &=
    \frac{1}{2}\sum_{m=1}^{M}\omega_{m}\bigl(x_{m}^{2} + p_{m}^{2}\bigr)
    +
    \sum_{i=1}^{N}\sum_{\alpha}\varepsilon_{i} \mathcal{R}_{i}^{ z\alpha} s_{i}^{\alpha}
    +
    \sum_{i=1}^{N}\sum_{m=1}^{M}
    \Bigl[\sqrt{2} g_{i m} -\omega_{m} \lambda_{i m}\Bigr] s_{i}^{\mu} x_{m}
    \nonumber\\
    &\quad
    +
    \sum_{i,j=1}^{N}
    \left[\sum_{m=1}^{M} \left( \frac{\omega_{m}}{2} \lambda_{i m} \lambda_{j m} - \sqrt{2} g_{i m} \lambda_{j m} \right) \right]
    s_{i}^{\mu} s_{j}^{\mu}
    -
    \frac{1}{2}\sum_{i,j=1}^{N}\sum_{\gamma,\alpha,\beta}J_{i j}^{\gamma} \mathcal{R}_{i}^{ \gamma\alpha} \mathcal{R}_{j}^{ \gamma\beta}
    s_{i}^{\alpha} s_{j}^{\beta}.
\end{align}
\end{widetext}

Averaging over $\ket{\psi_{\rm b}}$, we obtain the effective spin Hamiltonian given by~\eqref{eq:Heff}, whose coefficients are given by: effective local fields, 
\begin{align}
    h_{i,\alpha}^{\rm eff} &(\Delta_R, \Gamma, \{\lambda\}) =
    \varepsilon_{i} \bigl\langle \mathcal{R}_{i}^{ z\alpha}\bigr\rangle_{\rm b}
    \nonumber\\
    &+ \delta_{\alpha,\mu} \sum_{m=1}^{M}
    \Bigl[\sqrt{2} g_{i m} - \omega_{m} \lambda_{i m}\Bigr] \Delta_{x_m},
\end{align}
arising from rotated on-site terms, spin-boson couplings, and a backaction coming from the rotated bare bosonic term;
and effective spin-spin interactions
\begin{align}
    J_{i j,\alpha \beta}^{\rm eff} (& \Delta_R, \Gamma, \{\lambda\})
    = \delta_{\alpha,\mu} \delta_{\beta,\mu} \sum_{m=1}^{M} 
    [ \frac{\omega_{m}}{2} \lambda_{i m} \lambda_{j m}
    \nonumber\\
    &- \sqrt{2} g_{i m} \lambda_{j m} ]
    - \frac{1}{2}\sum_{\gamma\in\{x,y,z\}} 
    J_{ij}^{\gamma} \bigl\langle \mathcal{R}_{i}^{ \gamma\alpha} \mathcal{R}_{j}^{\gamma\beta}\bigr\rangle_{\rm b}.
\end{align}
The first term represents an all-to-all spin-spin interaction due to virtual exchange processes and backaction from the spin-boson coupling, and the last term stands for the dressed bare spin-spin interactions.

Finally, the variational energy functional reads:
\begin{align}
    \mathcal{E} (\Delta_R, \Gamma, \{\lambda\}) &= \bra{\phi_{\rm s}} H_{\rm eff} \ket{\phi_{\rm s}}
    \nonumber \\
    &= E_b + \sum_{i=1}^{N}\sum_{\alpha\in\{x,y,z\}}
    h_{i\alpha}^{\rm eff}   \langle s_{i}^{\alpha} \rangle_{\rm s}
    \nonumber\\
    &+
    \sum_{i,j=1}^{N} \sum_{\alpha,\beta\in\{x,y,z\}}
    J_{i j,\alpha \beta}^{\rm eff}   \langle s_{i}^{\alpha}   s_{j}^{\beta} \rangle_{\rm s}.
\end{align}

\section{Expectation values on the non-Gaussian state}

To efficiently evaluate the averages on the non-Gaussian ansatz given by
\begin{equation}
    \langle O \rangle_{\rm NGS} = \bra{0,\phi} U_{\rm GS}^\dagger U_\lambda^\dagger O U_\lambda U_{\rm GS} \ket{0,\phi},
\end{equation}
where $\ket{0,\phi} = \ket{0_{\rm b}} \otimes \ket{\phi_{\rm s}}$, one computes $\Tilde{O}=U_\lambda^\dagger O U_\lambda$ and applies Wick's theorem to analytically evaluate $\langle \Tilde{O} \rangle_{\rm GS} = \bra{0_{\rm b}} U_{\rm GS}^\dagger \Tilde{O} U_{\rm GS} \ket{0_{\rm b}}$ explicitly in terms of variational parameters and spin operators. Finally, averaging the final equation in $\ket{\phi_{\rm s}}$ leads to $\langle O \rangle_{\rm NGS}$.
Below we derive and list explicit formulas for three central classes of observables.

Using $n = \frac{1}{2}(x^2 + p^2 - 1)$, we find
\begin{align}
    \langle n \rangle_{\rm NGS}
    &= \frac{1}{2} \bigg[ {\rm Tr}\{\Gamma\} + \Delta_x^2 + \Delta_p^2
    - 2 \Delta_x \sum_{\alpha} \lambda^\alpha \langle S^\alpha \rangle_{\rm s}
    \nonumber\\
    &+ \sum_{\alpha,\beta} \lambda^\alpha \lambda^\beta \langle S^\alpha S^\beta \rangle_{\rm s}
    - 1 \bigg],
\end{align}
where we used $\langle \mathcal{R}_{i}^{x\alpha} x_{m} \rangle_{\rm b} = \langle \mathcal{R}_{i}^{x\alpha} \rangle_{\rm b} \Delta_{m}^{x} + \mathrm{Cov} (\mathcal{R}_{i}^{x\alpha},x_{m})$, and since $\mathcal{R}$ is a function of $p$ only, then $\mathrm{Cov}(\mathcal{R},x)=0$ precisely when $x$ and $p$ are uncorrelated i.e., when $\Gamma$ is diagonal. For the simplified homogeneous single-mode scenario studied in this paper, the photon number reads:
\begin{equation}
    \langle n \rangle_{\rm NGS} = \sinh^2(\xi) + \frac{4g^2}{N\omega^2} (1-\lambda^2) \langle S^x \rangle_{\rm s}^2 + \frac{4 g^2}{N\omega^2} \lambda^2 \langle (S^x)^2 \rangle_{\rm s} ,
\end{equation}
where $\Delta_x$ was analytically minimized and eliminated.

The spin operator rotates as $s_i^\alpha \to \sum_\beta \mathcal{R}^{\alpha\beta}(p) s_i^\beta$, yielding
\begin{equation}
    \langle s_i^\alpha \rangle_{\rm NGS}
    = \sum_{\beta} \langle \mathcal{R}^{\alpha\beta}(p) \rangle_{\rm GS} \cdot \langle s_i^\beta \rangle_{\rm s}.
\end{equation}
The longitudinal magnetization shown in the main text follows
\begin{equation}
    \langle S^z \rangle_{\rm NGS} = \exp\left( - \frac{2\lambda^2 g^2}{N\omega^2} e^{-2\xi} \right) \langle S^z \rangle_{\rm s}.
\end{equation}

Products $s_i^\alpha s_j^\beta$ transform bilinearly under dressing, resulting in
\begin{equation}
    \langle s_i^\alpha s_j^\beta \rangle_{\rm NGS}
    = \sum_{\mu,\nu}
    \langle \mathcal{R}^{\alpha\mu}(p)  \mathcal{R}^{\beta\nu}(p) \rangle_{\rm GS}
    \cdot \langle s_i^\mu s_j^\nu \rangle_{\rm s}.
\end{equation}
Setting $\lambda_{j1}^{\alpha}=\lambda (g'/\omega) \delta_{\alpha x}$ one can obtain the effective spin-spin terms shown in the main text. With this choice, the expectations with double $\mathcal{R}$ can be reduced in terms of single ones, as $\mathcal{R}$ becomes a simple rotation matrix about $x$-axis from $SU(3)$.

\section{Systematic expansion of the variational manifold}\label{sec:appx_manifold}

Rather than treating the spin-boson coupling perturbatively, our approach incorporates a unitary entangling transformation into the Hamiltonian. The objective is to push the spin-boson correlations (entanglement) into the definition of the variational manifold, inspired by perturbative Schrieffer-Wolff transformations. In this section, we propose a systematic way to improve accuracy by increasing the variational manifold via a nested hierarchy.

The lowest-order generator, $G_1 = \lambda S^x p$, is inspired by the standard polaron decoupling. Applying the transformation $U_1 = \exp(i \lambda S^x p)$ to the Dicke Hamiltonian yields, to leading order:
\begin{align}
    H_1 &\approx \frac{\omega}{2} (x^2 + p^2) + \varepsilon S^z - \varepsilon \lambda S^y p + (g' - \omega \lambda) S^x x 
    \nonumber\\
    &+ \left( \frac{\omega \lambda^2}{2} - g' \lambda \right) (S^x)^2 - \frac{\varepsilon \lambda^2}{2} S^z p^2.
\end{align}

Standard perturbative decoupling formally eliminates the linear $S^x x$ interaction by fixing $\lambda = g'/\omega$. Under this condition, the leading-order Hamiltonian shifts the interaction to the momentum quadrature and higher-order terms
\begin{align}
    H_1 &\approx \frac{\omega}{2} (x^2 + p^2) + \varepsilon S^z - \frac{g'^2}{2\omega} (S^x)^2 
    \nonumber\\
    &- \varepsilon \frac{g'}{\omega} S^y p - \frac{\varepsilon g'^2}{2\omega^2} S^z p^2.
\end{align}

However, this fixed-parameter transformation relies on the perturbative assumption that the coupling is weak ($g' \ll \omega$). In the ultrastrong coupling regime, fixing $\lambda = g'/\omega$ forces an over-displacement that completely invalidates the low-order truncation. By elevating $\lambda$ to a variational parameter, the ansatz bounds the displacement. It finds the optimal partial-decoupling that minimizes the energy, remaining well-behaved even in non-perturbative regimes where strict fixed-parameter transformations break down.

To systematically improve the ansatz, we identify the dominant residual interaction terms in $H_1$ and define the next-order generator $G_2$ to target them. The transformation $U_1$ leaves a residual momentum coupling $\sim S^y p$. The natural extension is therefore $G_2 = \lambda S^x p + \lambda^{(2)} S^y x$. The rotated Hamiltonian $H_2 = U_2^\dagger H_{\rm Dicke} U_2$, expanded to second order in the transformation parameters, becomes
\begin{align}
    H_2 &\approx \frac{\omega}{2}(x^2+p^2) + \varepsilon S^z + (g' - \omega \lambda + \varepsilon \lambda^{(2)}) S^x x
    \nonumber\\
    &+ (\omega \lambda^{(2)} - \varepsilon \lambda) S^y p + \left[ \frac{\lambda}{2} (\varepsilon \lambda^{(2)} - \omega \lambda) - g' \lambda \right] (S^x)^2 
    \nonumber\\
    &+ \left[ \frac{\lambda^{(2)}}{2} (\varepsilon \lambda^{(2)} - \omega \lambda) - g' \lambda^{(2)} \right] S^z x^2 
    \nonumber\\
    &- \frac{\lambda^{(2)}}{2} (\omega \lambda^{(2)} - \varepsilon \lambda) (S^y)^2 
    \nonumber\\
    &- \frac{\lambda}{2} (\omega \lambda^{(2)} - \varepsilon \lambda) S^z p^2.
\end{align}
Imposing the cancellation of both linear spin-boson interactions requires $\lambda = g' \omega / (\omega^2 - \varepsilon^2)$ and $\lambda^{(2)} = g' \varepsilon / (\omega^2 - \varepsilon^2)$. Under these conditions, the leading-order Hamiltonian diagonalizes the linear couplings, shifting the interactions to higher-order dispersive terms:
\begin{align}
    H_2 &\approx \frac{\omega}{2}(x^2+p^2) + \varepsilon S^z 
    \nonumber\\
    &- \frac{g'^2 \omega}{2(\omega^2 - \varepsilon^2)} (S^x)^2 - \frac{g'^2 \varepsilon}{2(\omega^2 - \varepsilon^2)} S^z x^2.
\end{align}
Here, the limitations of exact perturbative decoupling become explicit: the transformation possesses resonant denominators that diverge when $\omega \to \varepsilon$. A variational treatment of the parameters in $G_2$ inherently regulates these unphysical divergences.

This hierarchical expansion of the generator $G_n$ does not aim to analytically close the Lie algebra, but rather provides a targeted truncation of the operator basis that neutralizes the lowest-order non-commuting interaction channels. In general, one defines the generator such that
\begin{equation}
    G_n = \sum_{k=1}^n \lambda^{(k)} O_k, \quad U_n = e^{i G_n},
\end{equation}
where the variational parameter $\lambda^{(k)}$ controls the contribution of the $k$-th-order operator. The $n$-th-order non-Gaussian ansatz belongs to the manifold
\begin{equation}
    \mathcal{M}_n = \{ U_n U_{\rm GS}\ket{0_{\rm b}} \otimes \ket{\phi_{\rm s}} \}.
\end{equation}

The operators $G_n$ are constructed such that the lower-order generator is recovered by setting $\lambda^{(n)} = 0$. This guarantees that our non-Gaussian manifolds are strictly nested
\begin{equation}
    \mathcal{M}_1 \subset \mathcal{M}_2 \subset \dots \subset \mathcal{M}_n.
\end{equation}

Let $E^{(n)} = \min_{\Psi \in \mathcal{M}_n} \langle \Psi| H |\Psi\rangle$ be the approximate ground state solution. Since we are minimizing over a larger set at order $n+1$, we immediately obtain the strict variational inequality
\begin{equation}
    E^{(n+1)} = \min_{\Psi \in \mathcal M_{n+1}} \langle H \rangle \le \min_{\Psi \in \mathcal M_n} \langle H \rangle = E^{(n)}.
\end{equation}

Enlarging the variational manifold along these physically motivated directions systematically bounds the ground state energy from above, ensuring monotonic improvement in accuracy.

At each increased order, the complexity of the resulting effective Hamiltonian increases, and the associated variational parameter space grows, making such high-order treatments numerically intractable. In the present work, we restrict our attention to the leading-order form $G_1=\lambda S^x p$, which offers an optimal compromise between computational tractability and the ability to capture many-body correlations beyond mean-field. The accuracy of this truncation is verified through direct comparison with fully converged DMRG calculations in which the photon Hilbert space is truncated at a finite, but converged, $n_{\max}$.

\section{Spin-Boson reference DMRG truncation}\label{sec:appx_dmrg}

\begin{figure}[h]
    \centering
    \includegraphics[width=\columnwidth]{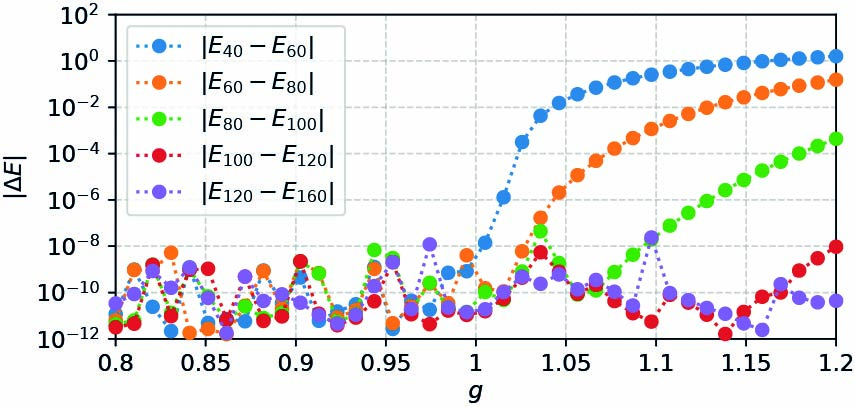}
    \caption{Convergence of the boson truncation for reference spin-boson DMRG in the Dicke-Ising model with $N=40$ and $J=0.8$.}
    \label{fig:appx_dmrg}
\end{figure}

Spin-boson DMRG calculations, here called reference DMRG to contrast with our solver, requires an explicit truncation of the bosonic Hilbert space. Restricting to $n_{\rm max}$ number of photons, the solver might converge towards an energy minima that does not fully capture the exact ground state. Therefore, a careful truncation analysis in employed to check whether the fixed value $n_{\rm max}$ is sufficient or not. We show in Fig.~\ref{fig:appx_dmrg} the energy difference for increasing $n_{\rm max}$ as a function of spin-photon coupling $g$ for the Dicke-Ising model with $N=40$ and $J=0.8$ to illustrate the numerical protocol. We note that a small $n_{\rm max} \sim \mathcal{O}(N)$ is sufficient to describe the system for $g<g_c$ while the necessary $n_{\rm max}$ grows with $g$ after the critical threshold, as photons condensate in the superradiant phase.  

\section{Expanded Benchmark Results}\label{sec:appx_benchmark}

In this section we expand the benchmark results to the four regimes discussed in the manuscript. Here, we focus on two main quantities, maximum bond dimension $\chi$ and energy error with respect to DMRG to represent cost and accuracy, respectively. 

\begin{figure}[t]
    \centering
    \includegraphics[width=0.95\columnwidth]{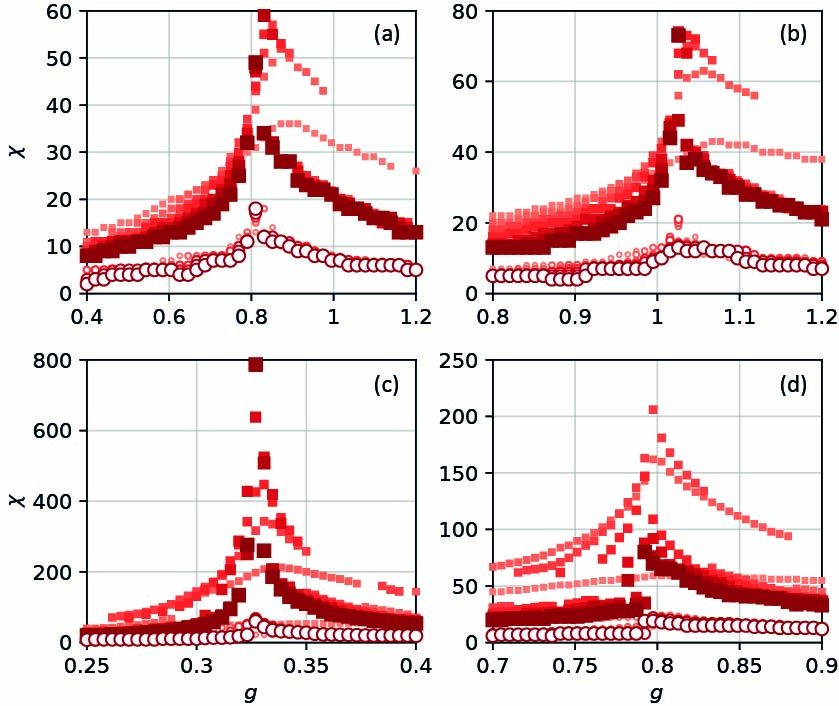}
    \caption{Maximum bond dimension of both NGS-DMRG and spin-photon DMRG using the Dicke-Ising model as a representative case. For the reference DMRG baseline, a converged photon cutoff $n_{\max}$ is fixed, which varies with $N$ and $J$. (a) $J=0.4$, (b) $J=0.8$, (c) $J=-0.3$, and (d) $J=-0.8$. Marker sizes scale proportionally with $N$.}
    \label{fig:app_costacc_1}
\end{figure}

\begin{figure}[h]
    \centering
    \includegraphics[width=0.95\columnwidth]{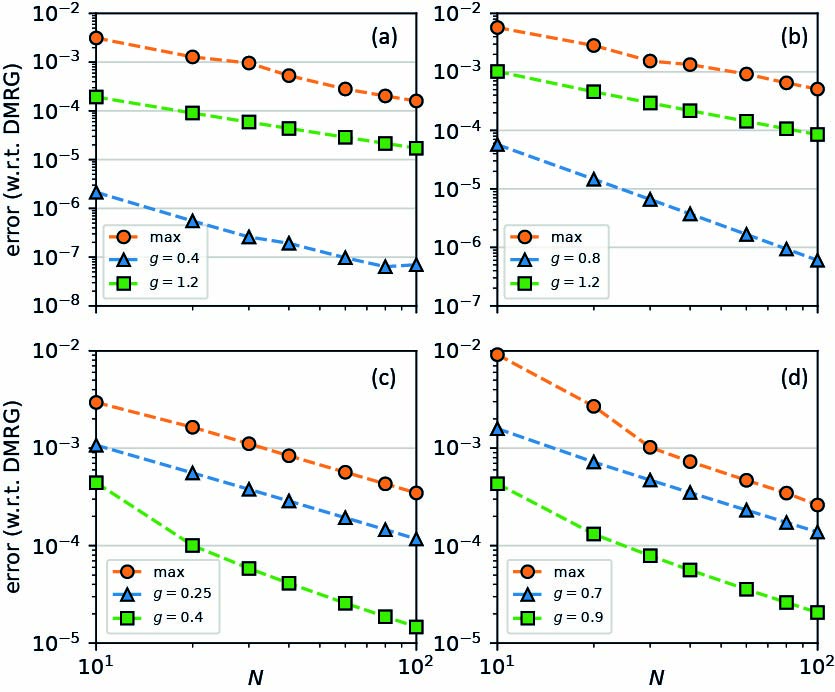}
    \caption{Energy error of the NGS-DMRG method with respect to reference DMRG, $|E_0-E_0^{\rm ref}|$, as a function of $N$ using the Dicke-Ising model as a representative case. For the reference DMRG baseline, a converged photon cutoff $n_{\max}$ is fixed, which varies with $N$ and $J$. (a) $J=0.4$, (b) $J=0.8$, (c) $J=-0.3$, and (d) $J=-0.8$.}
    \label{fig:app_costacc_2}
\end{figure}

We show in Fig.~\ref{fig:app_costacc_1}, that
in the ferromagnetic regime with $-\varepsilon/4 < J < \varepsilon/2$, the maximum bond dimension from both solver solutions remains small throughout the entire $g$ range shown, peaking around $g_c$ and decaying towards both normal and superradiant phases.
After the tri-critical point, $J > \varepsilon/2$, the ferromagnetic to superradiant QPT acquires a first-order nature, increasing numerical overhead. However, NGS-DMRG produces comparatively similar results, with the bond dimension still mostly below 20 and a similar energy difference. In the antiferromagnetic regime where $J < -\varepsilon/4$, as strong competition between Ising and Dicke interactions, the numerical cost increases by an order of magnitude and finite-size effects are become prominent, resulting in larger bond dimension for some small sizes. Interestingly, within the narrow intermediate region, the bond dimension of both solvers is comparatively small at large sizes, leading to a comparatively similar cost. However, DMRG presents a large steep peak around the critical threshold that grows with $N$ within the analyzed values. 
NGS-DMRG accuracy is well-controlled and decays roughly exponentially with $N$ throughout all studied regimes, as shown in Fig.~\ref{fig:app_costacc_2}.

\end{document}